\documentclass[multphys,vecphys]{svmult}

\usepackage{natbib}         
\usepackage{graphicx}        
\usepackage{multicol}        
\usepackage[bottom]{footmisc}

\def\deg{$^\circ$}
\newcommand{\smallmbox}[1]{\mbox{\footnotesize{#1}}}

\begin{document}

\title*{AMBER on the VLTI: data processing and calibration issues}
\titlerunning{AMBER/VLTI data calibration}
\author{Florentin Millour\inst{1}\and
  Romain Petrov\inst{2}\and
  Fabien Malbet\inst{3}\and
  Eric Tatulli\inst{4}\and
  Gilles Duvert\inst{3}\and
  G\'erard Zins\inst{3}\and
  Evelyne Altariba\inst{3}\and
  Martin Vannier\inst{5}\and
  Oscar Hernandez\inst{3}  \and
  Gianluca Li Causi\inst{6}}
\authorrunning{F. Millour et al.}
\institute{Max-Planck-Institut f\"ur Radioastronomie,
  \texttt{fmillour@mpifr-bonn.mpg.de}
  \and Laboratoire Universitaire d'Astrophysique de Nice
  \and Laboratoire d'AstrOphysique de Grenoble
  \and Observatorio di Arcetri
  \and European Southern Observatory
  \and Rome Astronomical Observatory
}
%
%
\maketitle

\section{Introduction}
\label{sec:intro}

We present here the current performances of the AMBER / VLTI
instrument for standard use and compare these with the offered modes
of the instrument. We show that the instrument is able to reach its
specified precision only for medium and high spectral resolution
modes, differential observables and bright objects.

For absolute observables, the current achievable accuracy is strongly
limited by the vibrations of the Unit Telescopes, and also by the
observing procedure which does not take into account the night-long
transfer function monitoring.

For low-resolution mode, the current limitation is more in the data
reduction side, since several effects negligible at medium spectral
resolution are not taken into account in the current pipeline.

Finally, for faint objects (SNR around 1 per spectral channel),
electromagnetic interferences in the VLTI interferometric laboratory
with the detector electronics prevents currently to get unbiased
measurements. Ideas are under study to correct in the data processing
side this effect, but a hardware fix should be investigated seriously
since it limits seriously the effective limiting magnitude of the
instrument.

\section{The AMBER instrument}
\label{sec:amber}

\subsection{Short description}
\begin{figure}[tbp]
  \centering
  \includegraphics[height=\textwidth, angle=-90]{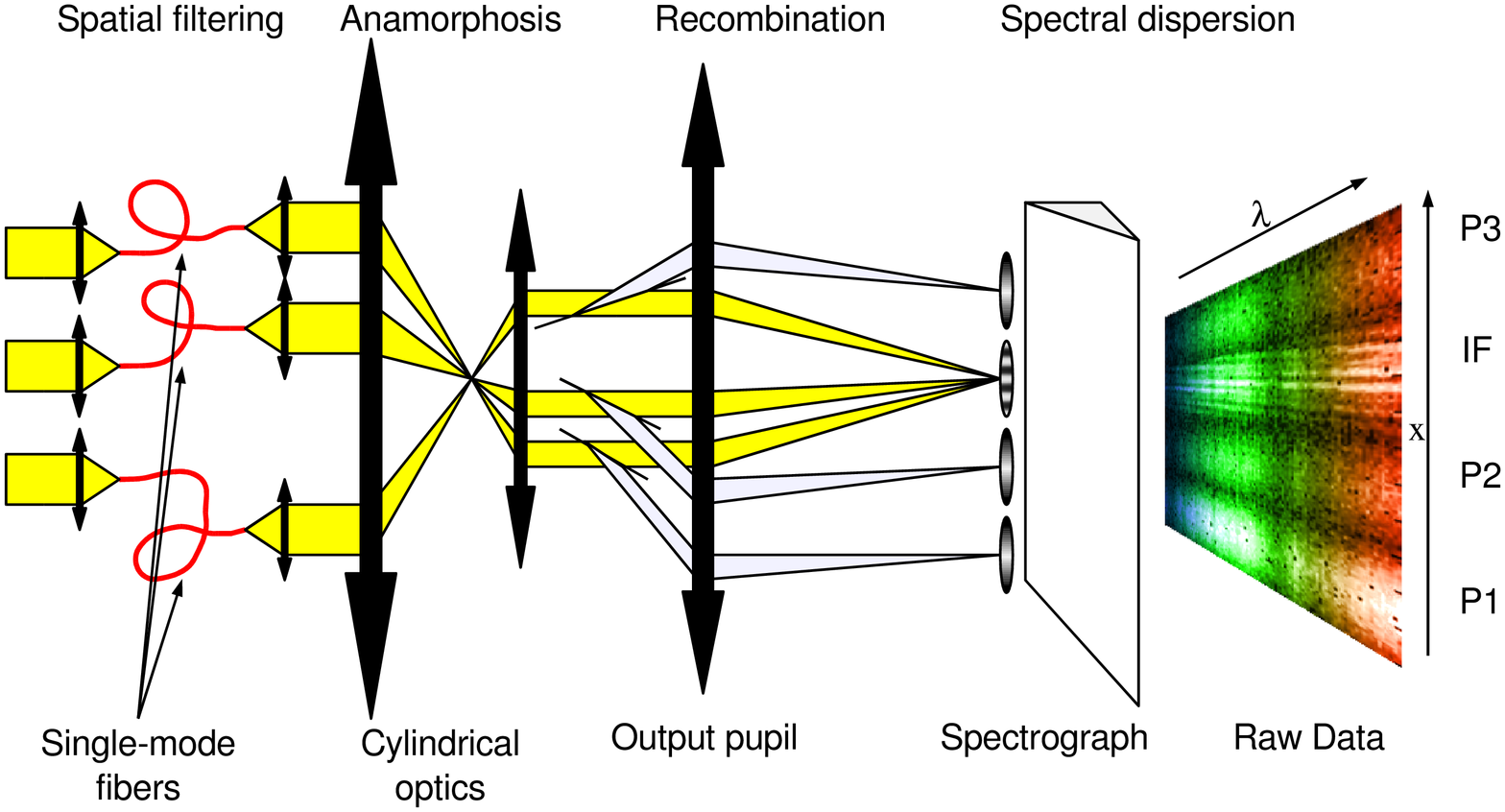}
  \caption{The AMBER optical schematics, showing the principal
    elements of the instrument: the spatial filtering is made with
    optical fibers, then an anamorphosis optics shrinks the beam in
    one direction to feed it into a long slit spectrograph where the
    spectrally dispersed fringes are finally imaged on the detector.}
  \label{fig:amberSchema}       
\end{figure}

AMBER is the near-infrared interferometric re-combiner of the VLTI. Its
general and technical descriptions are held in the two articles
\citet{2007A&A...464....1P, 2007A&A...464...13R}. In short, it features
simultaneous observations in $J$, $H$ and $K$ bands observations, low
(R=35), moderate (R=1500) and high (R=12000) spectral resolutions, and
3 telescopes operation. The use of optical fibers to improve
calibration and the multiaxial scheme adopted complete this short view
of the instrument (see Fig. \ref{fig:amberSchema}). The AMBER
instrument features also a limited number of pixels in the fringes
leading to the use of very specific algorithms for data reduction.

\subsection{AMBER data processing: principle}

The AMBER data processing is based on the fitting in the image plane
of the fringe pattern. One can find a complete description of the
process in the article \citet{2007A&A...464...29T}. The resulting
basic information is a measurement of the coherent flux (instantaneous
complex coherence factor multiplied by the flux) for each single
frame. Three time-averaged squared visibility, a closure phase and
three differential phases can be extracted from these measurements,
using respectively specific techniques such as quadratic estimator
\citep{2003A&A...400.1173P}, bi-spectrum estimator
\citep{1990SPIE.1351..522H} and inter-spectrum estimator
\citep{1981LowOB...9..165B}. Added to these interferometric
observables, AMBER provides the object spectrum which adds
simultaneous velocimetric measurements to the purely geometric
measurements brought by the interferometric observables.

\section{AMBER and the VLTI}

\subsection{How AMBER behaves on VLTI ?}
\label{sect:AMBERVLTI}

\begin{figure}[tbp]
  \centering
  \begin{tabular}{cc}
    \includegraphics[width=0.48\textwidth]{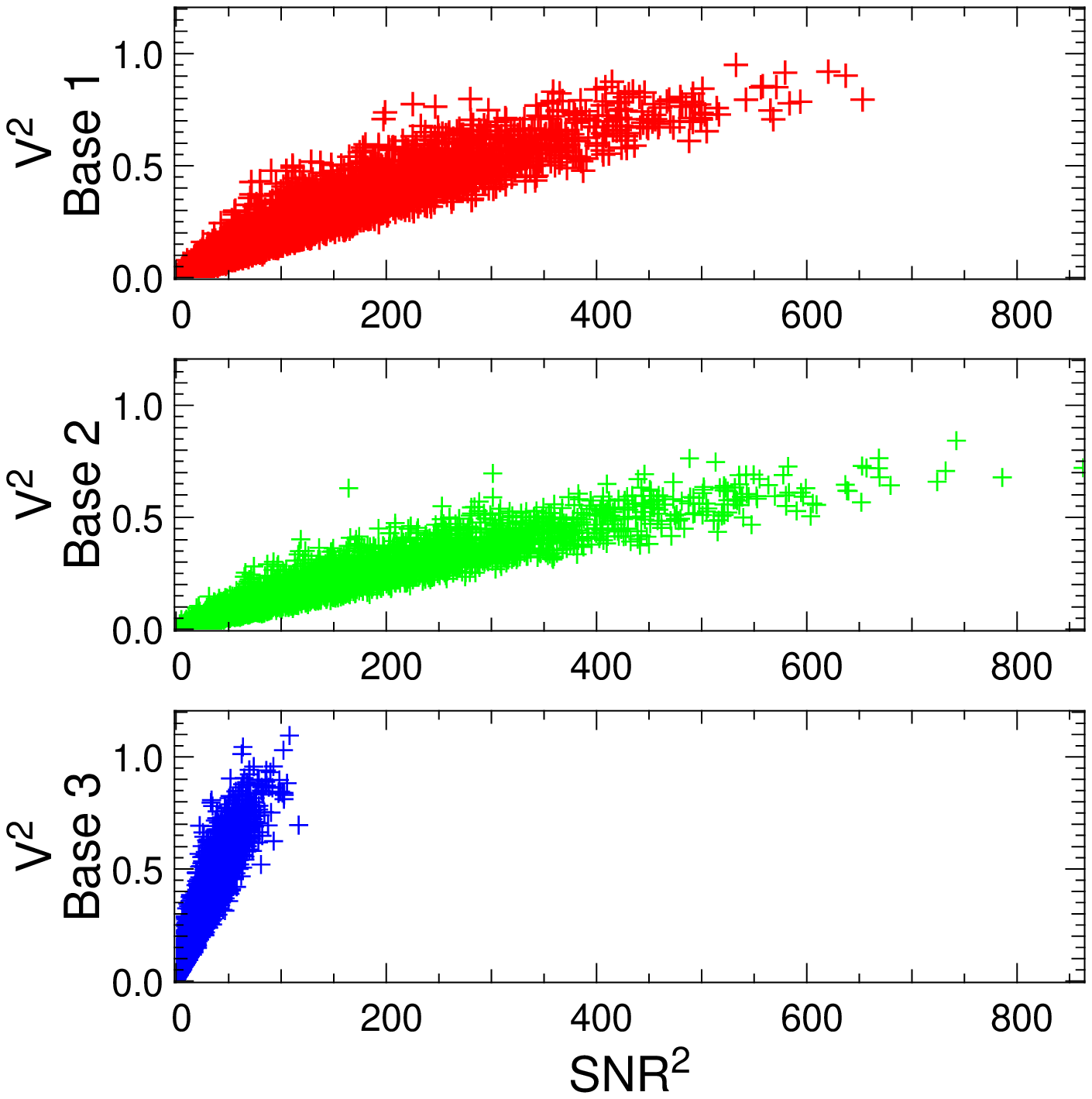}&
    \includegraphics[width=0.48\textwidth]{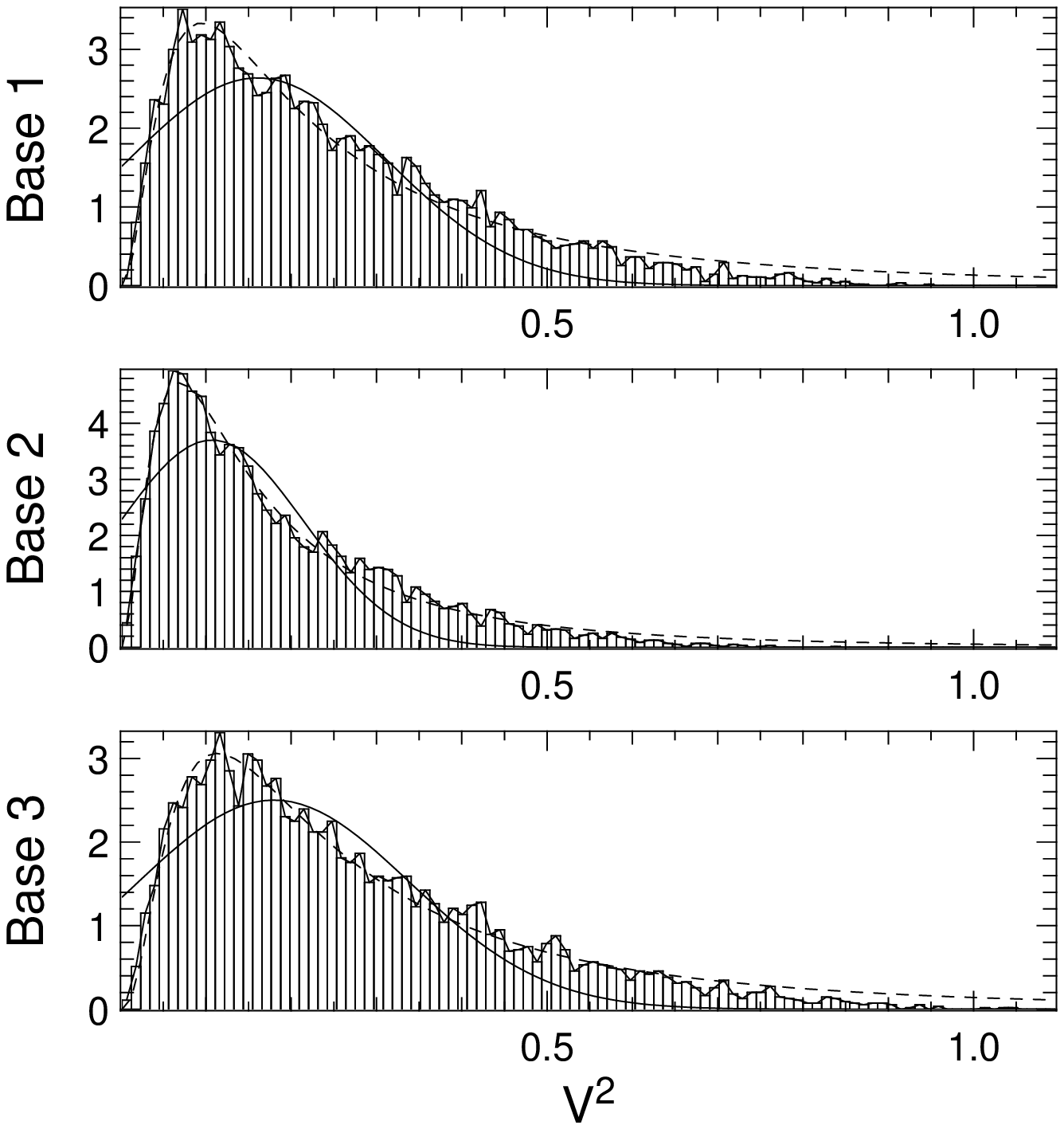}
  \end{tabular}
  \caption{\emph{Left:} A correlation plot between fringe SNR and
    visibilities shows a strong correlation, linked to the
    domination of the jitter effect (blurring of the fringes by their
    move during integration time).
    \protect\\\emph{Right:} AMBER squared visibilities histogram on the
    same star, showing the highly non-symmetric effect induced by UT
    vibrations. The resulting histogram looks like a log-normal
    distribution (dashed line) and not like a Gaussian distribution (solid
    line), which makes it difficult to extract an average and an error.}
  \label{fig:HistogVisi}       
\end{figure}

The AMBER instrument was installed and tested at the Paranal
observatory (ESO, Chile) during the month of may 2004
\citep{2004SPIE.5491.1089R}. Since then, a series of commissioning
were performed to check the performances of the AMBER instrument
together with the VLTI infrastructure. These first tests showed that
the UTs have vibrations that affects strongly the AMBER signal (see
Fig.~\ref{fig:HistogVisi}) with a drastic average instrumental
visibility decrease with regards to specifications (20\% instead of
80\%, leading to a loss of a factor 4 in signal to noise ratio). These
vibrations comes mainly from the Coud\'e train of the UTs and are in
the process of being damped by ESO. However, one has to consider using
the AMBER instrument in the today limited state of VLTI, taking into
account these vibrations.

Therefore, changes in observing and data processing strategies were
needed for AMBER to successfully be opened to the community:

\begin{itemize}
\item A longer exposure time (i.e. more individual short exposure
  frames) is needed to maximize the chances to get some ``useful''
  frames for the data processing.
\item A frame-selection process, i.e. a removal of spurious frames
  where there is either no fringes, no flux or a too high piston has
  been added to the data processing software.
\end{itemize}

\begin{figure}[tbp]
  \centering
  \includegraphics[width=0.96\textwidth, angle=0]{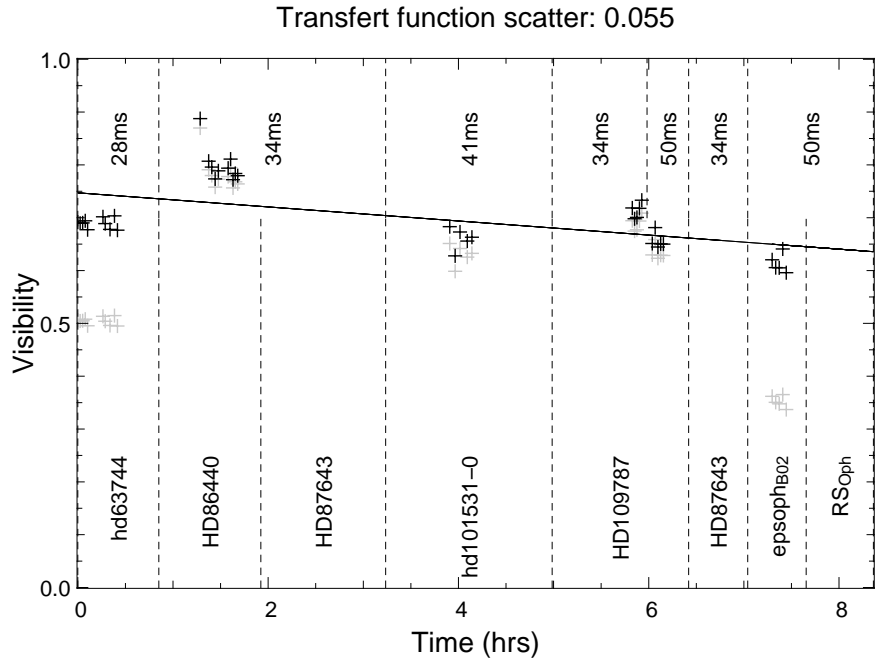}
  \caption{Instrument+atmosphere transfer function during the night of
    17/02/2006. In gray are the original visibilities and in black the
    visibilities corrected from the intrinsic calibration star's
    visibilities. The stars where no visibilities are plotted are the
    science stars. One can see the frame exposure time dependence of the
    transfer function for the star HD 109787.}
  \label{fig:transfunc}       
\end{figure}

\subsection{The standard operating mode performances}

These observing and data processing strategies are used today on the
AMBER/VLTI instrument, allowing a better visibilities histogram
(closer to a  Gaussian and therefore easier to compute realistic error
bars), but are quite observing time-consuming (about 50 to 90\% of the
shutter-opened observing time is lost in practice, added to the
already large overheads linked to optical interferometry). This allows
an internal precision (i.e. visibilities scatter inside an exposure
time) of roughly 0.01 to 0.05 for a bright star ($K\leq4$ at medium
spectral resolution), depending on the observing conditions (wind,
vibrations, seeing). However, in practice, the AMBER visibilities
precision cannot be better than 0.05 due to several strong limitations
coming from the infrastructure itself and from the observing strategy
used at Paranal.

\begin{itemize}
\item The time between two measurements cannot be less than 30\,min
  (for star and calibrators, i.e. 1h between 2 science measurements),
  leading to very large time gaps in the
  instrumental/atmospheric/vibrations transfer function (see
  Fig.~\ref{fig:transfunc}). This prevents today from interpolating
  such transfer function and gain in precision and stability of the
  measurement.
\item The individual frames exposure times changes a lot during the
  night (see Fig.~\ref{fig:transfunc}), leading to miscomparisons
  between full-night data sets, which provokes a typical night-long
  visibilities scatter of 0.05. This translates into typical
  calibrated visibilities errors of about 0.07.
\end{itemize}

\begin{table}
  \centering
  \caption{Error bars order of magnitude one can expect from the AMBER
    instrument in the current status using UTs for calibrated
    measurements. Visibility errors are dominated by the transfer function
    error and not by internal visibilities scatter. LR differential data
    reduction is highly biased by the atmospheric phase bias and the
    errors given here take into account this bias as an error. The
    figures given here can be seen as the result of a hard work data
    processing and very careful calibration process and not as a
    pipeline black-box output error estimate.}
  \label{tab:AMBER_today_performances}       
  %
  %
  \begin{tabular}{|l|c|c|}
    \hline
    \hline
    & LR (R=35)  & MR (R=1500)  \\
    Observable & Bright star ($K=5$) & Bright star ($K=3.5$) \\
    \hline
    $V$        & 0.07 & 0.07 \\
    $V_{\smallmbox{diff}}$    & 0.1 & 0.01 \\
    $\phi_{\smallmbox{diff}}$ (rad.) & 0.1 & 0.01 \\
    $\psi_{\smallmbox{123}}$ (rad.)  &  0.01 & 0.05 \\
    \hline
    \hline
  \end{tabular}
\end{table}

The table \ref{tab:AMBER_today_performances} summarizes the current
situation with typical observing conditions and the AMBER/VLTI
instrument. Please note that this is an indicative table and does not
represent all the conditions, which can change strongly depending on
the air-mass, seeing, vibrations conditions, etc. Therefore,
improvements can be expected for AMBER in standard mode, following
several tracks for improving operation:

\begin{itemize}
\item Restrain strongly the number of available and effectively used
  exposure times during the observing nights. This would probably
  enhance the final transfer function scatter and therefore the
  calibration accuracy.
\item Accelerate the time between observations by working on the
  telescopes overheads: from the beginning of AMBER operation to
  today, huge improvements have been achieved, and the arrival of
  FINITO will improve again these overheads.
\item Improve the data processing software, in terms of accuracy, so it
  take into account the identified problems and proposed solutions. An
  effort is also needed in terms of ergonomy and documentation.
\end{itemize}

\subsection{Closure phase and differential phases}

\begin{figure}[tbp]
  \centering
  \includegraphics[width=0.96\textwidth, angle=0]{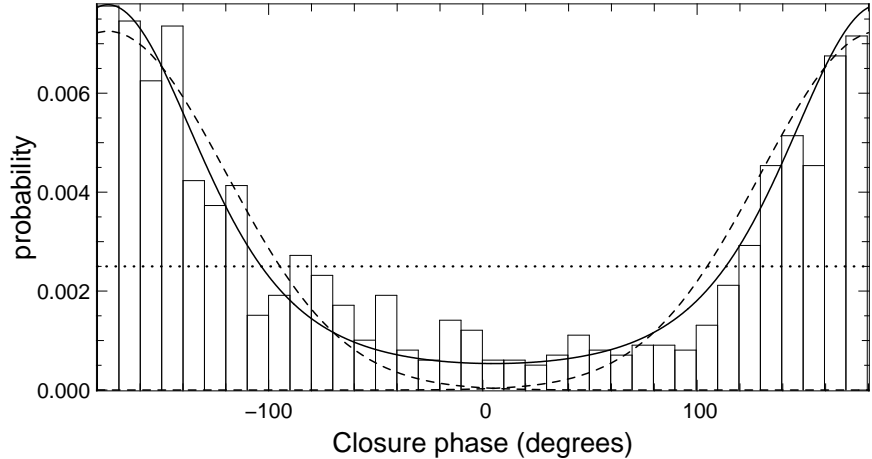}
  \caption{Example of a closure phase histogram on an observed star
    ($\epsilon$ Sco, courtesy of O. Chesneau), illustrating the difference
    between a Gaussian distribution (dashed line) and a wrapped phase
    distribution (solid line): the phase distribution is in-between a
    Gaussian-like (dashed line) and a white noise distribution (dotted
    line), leading to a problematic estimation of the error bars.}
  \label{fig:histogramClos}       
\end{figure}

Due to the low number of frames where all three fringes patterns from
the three baselines are present together, the closure phases are very
much affected by the current state of the VLTI. Therefore, the best
achievable closure phase accuracy on a bright star is of the order of
10$^{-2}$ radians, i.e. $\sim$1\deg, in low spectral resolution, and
about 10$^{-1}$ radians, i.e. $\sim$10\deg\ in medium spectral
resolution. For ``standard'' applications, this is in general
sufficient, but for high dynamics or high accuracy measurements, this
low precision is very much killing the use of such observable. Added
to that, the specific behaviour of phases obliges one to compute very
specifically the error bars (see Fig.~\ref{fig:histogramClos}),
resulting in under-evaluated error bars in the current data processing
software and very noisy data (but this will be solved in the next
releases).

The situation on differential phases is much better, with already
reached 10$^{-2}$ radians accuracy at medium spectral resolution
\citep{2007A&A...464...59M} and 10$^{-3}$ radians accuracy (but with a
10$^{-1}$ radians bias) at low spectral resolution
\citep{2006dies.conf..291M}. However, systematic biases related to the
amount of water vapour are still under investigation and the
calibration of such biases for low spectral resolution is still under
development.

\subsection{The low flux issue}

\begin{figure}[tbp]
  \centering
  \includegraphics[width=0.96\textwidth,
  angle=0]{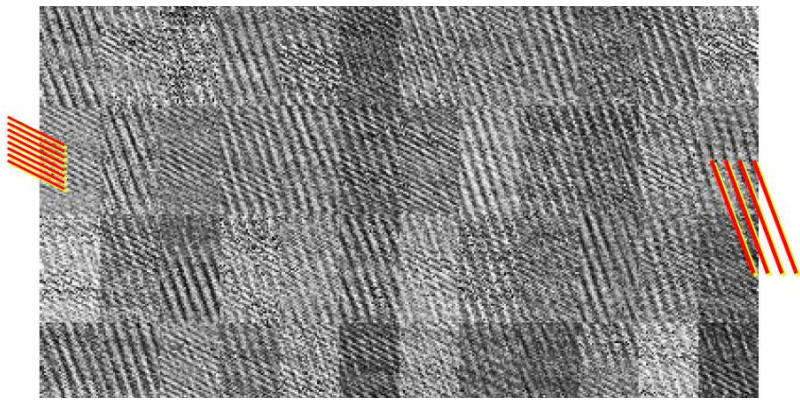}
  \caption{AMBER detector fringes induced by electromagnetic
    interferences \citep{Li2007}.}
  \label{fig:detFringes}       
\end{figure}

For low number of photons, the AMBER signal switches from a
photon-noise driven behaviour to a detector-noise driven one. In case
the detector behaviour is well known, this signal with few photons is
workable with a lower signal-to-noise regime than with high
flux. However, for AMBER, the detector noise behaviour has changed
between the integration laboratory in Grenoble in 2003 and the Paranal
interferometric laboratory where it is installed today:
electromagnetic interferences from an unknown source occurs and
creates a correlated noise which appears as ``detector
fringes'' (see Fig.~\ref{fig:detFringes}). Therefore in the current
data reduction scheme and knowledge of the detector put in the AMBER
software, low flux data reduction results in non reliable results.

Therefore, while ESO puts manpower to solve this problem in the
hardware side \citep{Mardones2007}, this problem is also under
investigations in the data reduction software side \citep{Li2007}, in
order to achieve a workable data reduction solution for already
observed targets. The idea is there to evaluate the detector
correlated noise from exposures without fringes (dark or photometric
beams) and to try to subtract it from the interferometric beam where
the pattern affects the fringes. A prototype version of this algorithm
can be found on the AMDC\footnote{AMBER detector cleaner,
  \texttt{http://www.mporzio.astro.it/$\sim$licausi/AMDC/}} web page.

However, this study is only for the already-acquired data and this
problem affects also the on-site real time acquisition of the targets,
preventing AMBER to reach its goal limiting magnitude without the
external fringe tracker FINITO. 

\subsection{The low resolution issues}

\begin{figure}[tbp]
  \centering
  \includegraphics[width=0.96\textwidth,
  angle=0]{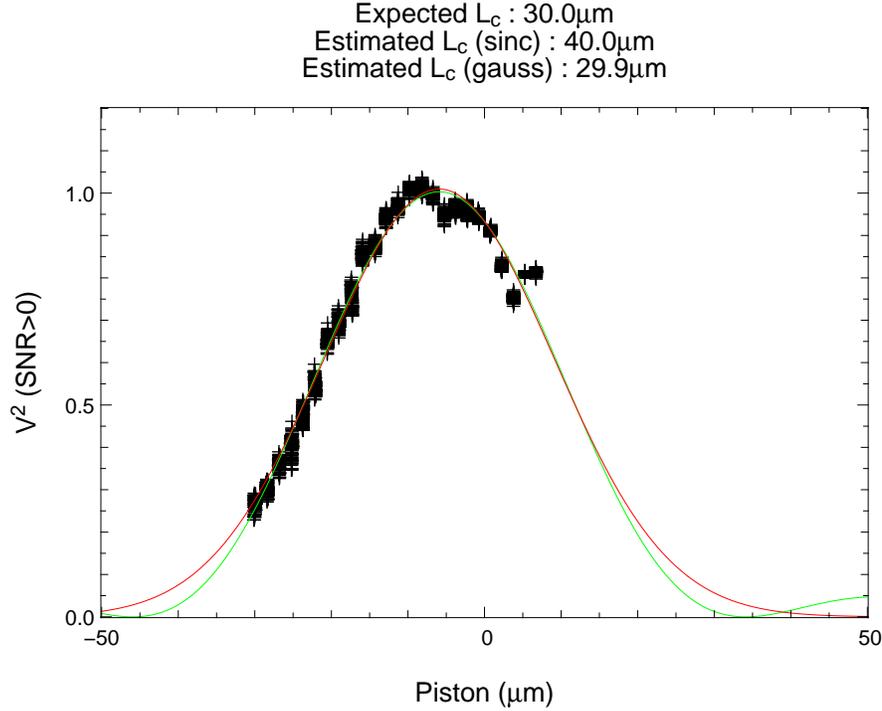}
  \caption{Coherence length (L$_c$) dependence of the visibilities of
    AMBER measured in lab. using the internal piezo mirrors of the
    instrument. It corresponds to a Gaussian of FWHM L$_c$.}
  \label{fig:envCoherence}       
\end{figure}

AMBER is facing a series of data processing issues very specific to
low resolution observations, and which prevents it from working in an
optimal way. We present here a series of identified points to enhance
significantly the data processing in this mode:

\begin{itemize}
\item The jitter effect affecting the visibilities
  (Fig.~\ref{fig:HistogVisi}) is very important in the way that it
  introduces an ``artificial'' slope to the visibilities, very hard to
  calibrate since this jitter effect cannot be calibrated efficiently
  in the current state of the infrastructure.
\item The coherence length visibility decrease plays also an important role
  (see Fig.~\ref{fig:envCoherence}), and is difficult to calibrate.
\item Problems of bias removal in the squared visibilities introduce a
  visibility flux dependence, which is highly problematic for low
  resolution, since between the center and the edges of a band, a
  difference of flux of up to 100 can be found.
\end{itemize}

All these issues except the first one have an associated solution,
included hopefully in the next version of the AMBER data reduction
software. The jitter issue is still under investigation and has not
reached up to now a level of comprehension allowing us to present a
solution. For this issue, the use of FINITO (available first with ATs
in P80 and probably soon after with the UTs) is expected to improve a
lot the problem.

\section{Conclusion}

We presented here an overview of what AMBER can do in the current
state of the instrument, data reduction pipeline and
infrastructure. The table \ref{tab:AMBER_today_performances} gives
typical figures (taken from the personal experience of the authors,
and that must be taken as indicative values and not as specified
performances) of what can reach in practice the AMBER instrument using
the Unit Telescopes of VLTI. With the use of Auxiliary Telescopes and
FINITO in a near future, these figures are expected to improve a lot.

As one can see, the instrument does not work in
optimal conditions and there is room for near future improvements as
well as long term instrumental study to correct all the expected and
unexpected effects on the AMBER signal. However, the huge potential of
the AMBER instrument has been already proved by the numerous first
articles published in a special feature of A\&A
\citep{2007A&A...464...43M,
  2007A&A...464...55T,
  2007A&A...464...59M,
  2007A&A...464...73M,
  2007A&A...464...87W, 
  2007A&A...464..107M,
  2007A&A...464..119C,
  Domiciano2007}, and
many other new successful observing programs can be expected for the
future, even with the infrastructure-limited performances of the
instrument.

\subsection*{Acknowledgements}
\footnotesize{

This paper is based on data taken at the Paranal observatory, ESO,
Chile, and during the AMBER integration in Grenoble in 2003.

The AMBER project (The structure and members of the AMBER Consortium
can be found in the
AMBER website\footnote{\texttt{http://amber.obs.ujf-grenoble.fr}}\label{note:AMBERwebsite}) has been
founded by the French \emph{Centre National de la Recherche
  Scientifique} (CNRS), the \emph{Max Planck Institute f\"ur
  Radioastronomie} (MPIfR) in Bonn, the \emph{Osservatorio Astrofisico
  di Arcetri} (OAA) in Firenze, the French Region \emph{Provence Alpes
  C\^ote D'Azur} and the \emph{European Southern Observatory} (ESO).


The AMBER data reduction software \texttt{amdlib} is freely available
on the AMBER website. It has been linked to the open source
software Yorick\footnote{\texttt{http://yorick.sourceforge.net}} to
provide the user friendly interface \texttt{ammyorick}.
}

%
\bibliographystyle{aa}
\bibliography{biblio}
%


\end{document}